\documentstyle[12pt]{article}
\begin{document}

\newcommand{\mpio}{\stackrel{\circ}{m}_\pi^2}
\newcommand{\mpi}{\stackrel{\circ}{m}_\pi}
\newcommand{\m}{\stackrel{\circ}{m}}
\newcommand{\mo}{\stackrel{\circ}{m}^2}
\newcommand{\moo}{\stackrel{\circ}{m}^4}
\newcommand{\Ido}{\stackrel{\circ}{I}_2}
\newcommand{\Ito}{\stackrel{\circ}{I}_3}
\newcommand{\Ico}{\stackrel{\circ}{I}_4}
\newcommand{\Ipo}{\stackrel{\circ}{I}_5}
\newcommand{\mpioo}{\stackrel{\circ}{m}_\pi^4}
\newcommand{\ho}{\stackrel{\circ}{h}_1}
\newcommand{\hoo}{\stackrel{\circ}{h}_2}
\newcommand{\G}{\stackrel{-}{G}}

\centerline{\Large\bf Low-energy dynamics of the $\gamma\gamma\rightarrow\pi\pi$
                      reaction}
\centerline{\Large\bf in the NJL model\footnote{Work supported in part by
CERN nos. - FAE/74/91 and - FIS/116/94, JNICT, CNPq, the Russian
Foundation for Basic Research No 94-02-03028 and the Ministry of Science
and Technology of the Republic of Slovenia.}}

\vspace{1.5cm}

\centerline{\large
     B. Bajc$^{\dag}$, A.H. Blin$^{\S}$, B. Hiller$^{\S}$, M.C. Nemes$^*$,}
\centerline{\large  A.A. Osipov$^{\ddag}$ and M. Rosina$^{\dag X}$}

\vspace{1.5cm}

\noindent
$^{\dag}$Institut Jo\v zef Stefan, Jamova 39, p.p.100, 61111 Ljubljana,
Slovenia

\vspace{0.5cm}

\noindent
$^{\S}$Centro de F\'{\i}sica Te\'{o}rica, Dpt. de F\'{\i}sica, P-3000 Coimbra,
Portugal

\vspace{0.5cm}

\noindent
$^*$Dpt. de F\'{\i}sica, ICEX, Universidade Federal de Minas Gerais,
C.P.702, 31270 Belo Horizonte-MG, Brasil

\vspace{0.5cm}

\noindent
$^{\ddag}$Joint Institute for Nuclear Research, Laboratory of Nuclear
Problems, 141980 Dubna, Moscow Region, Russia

\vspace{0.5cm}

\noindent
$^X$Dpt. of Physics, University of Ljubljana, Jadranska 19,
p.p. 64, 61111 Ljubljana, Slovenia

\vspace{3cm}

\centerline{ABSTRACT}

\vspace{1.0cm}

We calculate the one-quark-loop amplitude for the low energy $\gamma\gamma\to\pi
\pi$ collision in the context of the Nambu and Jona-Lasinio model with scalar
and pseudoscalar four quark couplings to all orders in the external momenta.
We show that the NJL predictions for the $\gamma\gamma\to\pi^+\pi^-$ reaction
are not far from the Born amplitude, which is close to the data, and is
compatible with the chiral perturbation theory estimations. We determine the
corrections given by the NJL model in leading order of $1/N_c$
to the chiral loop amplitude for $\gamma\gamma\to\pi^0\pi^0$.
Numerical results for the $\gamma\gamma\rightarrow\pi\pi$ cross sections
and for pion polarizabilities are given.

\vfill

\newpage
\section{Introduction}
\vspace{0.5cm}

Currently there is much interest in  calculations of the cross sections
for the $\gamma\gamma\rightarrow\pi^+\pi^-$ and
$\gamma\gamma\rightarrow\pi^0 \pi^0$ modes. In \cite{1} these processes
were considered in the framework of chiral perturbation theory (CHPT)
\cite{2}-\cite{4} in the next-to-leading order. Moderate enhancement for
the $\pi^+ \pi^-$ production near  threshold and a small steadily
rising cross section for the $\pi^0\pi^0$ production were shown. For the
charged channel the chiral calculation \cite{1} at the next-to-leading
order is in good agreement with experimental data \cite{5}. For the
neutral channel the theoretical picture is more complicated. Tree diagrams
are absent at the $p^4$ level for $\gamma\gamma\to \pi^0\pi^0$ which
starts with meson-one-loop graphs. The one-loop result \cite{1} for
$\pi^0\pi^0$ production deviates from the Crystal Ball data \cite{6} by
25-30\% in the amplitude. Only the two-loop calculations (the $p^6$ order)
\cite{7} have reconciled CHPT with experiment and with dispersion
calculations \cite{8}-\cite{10}. Three new constants $d_i$ which appear at
this order in the $L^{(6)}$ CHPT Lagrangian have been estimated with
resonance saturation. At present it is impossible to determine $d_i$ from
other processes. This still leaves an uncertainty in the CHPT calculation.

Formerly \cite{11} an attempt was made to estimate the tree-level
contribution which starts at order $p^6$ for $\pi^0\pi^0$ production.
Within specific models one can predict higher order constants of the
effective meson Lagrangian at least at the tree level, i.e. to leading
order in $1/N_c$. One can expect that  tree-level contributions
dominate over meson loops when they work at the same order in chiral
counting. In this case, calculation of these tree-level contributions
yields the main part of higher order corrections to the leading $p^4$
result for neutral pion production. Therefore there arises an opportunity
not only
to estimate the corrections appearing in the $p^6$ approximation but also
to advance in energy. This yields indirect information on the convergence
rate of the chiral series. Besides, it is not impossible that in some
cases strong compensations of contributions from box diagrams on the one
hand and from $\sigma$-exchange diagrams on the other can be observed in
the $p^6$ approximation. If so, the contributions of the order of $p^8$
will be particularly important.

In this paper, following in general the main idea of \cite{11} we
calculate the leading  $1/N_c$ contribution to $\gamma\gamma\to\pi\pi$
to {\it all orders in external momenta and quark masses}, in the
context of the Nambu and
Jona-Lasinio (NJL) model \cite{12}. A small number of
parameters is an obvious advantage of the model as compared with the
resonance saturation method where the number of parameters increases when
new transitions and higher chiral orders are considered. Mathematical tools
allowing the higher terms of the chiral expansion to be calculated are
provided by the momentum-space bosonization method \cite{13}. It is clear
that other known methods \cite{14}-\cite{fmd} would lead us to the same
results. We consider the NJL Lagrangian only with pseudoscalar and scalar
couplings. The vector and axial-vector contributions are beyond the scope
of this paper\footnote{We refer the readers to the papers \cite{13} and
\cite{17} where the $\pi\pi$-scattering amplitude has been calculated in
the same framework using the NJL Lagrangian without spin-1 mesons and with
them. The result clearly indicates only a small impact of a model extension
like this on the final picture near threshold.}. Also, the main bulk of
events are peaked near on-shell values of the photons, so we present
calculations only for real photons.

In the recently published papers \cite{bb} and \cite{bel} the NJL model
is also used to investigate the process $\gamma\gamma\to\pi^0\pi^0$.
In contrast to this, we consider not only the neutral mode but also
the charged
one $\gamma\gamma\to\pi^+\pi^-$. Besides, in the above papers the
amplitude was derived only to ${\cal O}(p^6)$ while we do not confine
ourselves to this approximation. While analyzing the referee's comments
on our work we learned of the results published in the preprint \cite{bfp}.
Its authors employed another method to take into account the total momentum
dependence of meson vertices \cite{fmd}. They managed to solve the problem
only numerically. In our paper we derive {\it analytical} expressions for
the amplitudes in question, with full momentum dependence,
and formulae for pion polarizabilities. This
allows us to construct chiral expansions of the amplitude and to investigate
the role of the first $p^6$ and $p^8$ approximations.

The article is organized as follows. In Sec. 2 we set up the notation.
In Sec. 3 we give a short description of the model and obtain
general expressions for amplitudes of the $\gamma\gamma\to\pi\pi$ reaction.
In Sec. 4 we consider the specific neutral mode $\gamma\gamma\to\pi^0\pi^0$,
obtain the cross section for this process and calculate the
polarizabilities of neutral pions. The same program is realized for the
charged mode in Sec. 5. Finally, a summary and concluding remarks are
presented in Sec. 6.

\vspace{0.5cm}
\section{Kinematics}
\vspace{0.5cm}

We consider the collision of two on-shell photons yielding two pions in
the exit channel
$$
  \gamma (p_1,\epsilon_{\mu})+\gamma (p_2,\epsilon_{\nu})\rightarrow\pi^a(p_3)
  +\pi^b(p_4),
$$
where $a,b$ are the isotopic indices. The matrix element for pion
production is
\begin{equation}
  T(p_1,p_2,p_3)=e^2\epsilon^\mu (p_1)\epsilon^\nu (p_2)T_{\mu\nu}(p_1,p_2,p_3),
\label{ampl}                                      
\end{equation}
where $e$ is the electric charge.

One can decompose $T_{\mu\nu}$ into Lorentz and parity invariant amplitudes
\begin{equation}
  T_{\mu\nu}=C_1g_{\mu \nu}+C_2p_{2\mu}p_{1\nu}+C_3p_{2\mu}p_{3\nu}+
  C_4p_{3\mu}p_{1\nu}+C_5p_{3\mu}p_{3\nu}
\label{a10}                                       
\end{equation}
where the terms proportional to $p_{1\mu}$ and $p_{2\nu}$ are not
considered since they drop out when contracted with the polarization
vectors in
(\ref{ampl}),
\begin{equation}
  p_1^{\mu}\epsilon_{\mu}(p_1)=p_2^{\nu}\epsilon_{\nu}(p_2)=0.
\label{a9}                                        
\end{equation}
From the Ward identities
\begin{equation}
  p_1^\mu T_{\mu \nu}=T_{\mu \nu}p_2^\nu =0,
\label{a11}                                       
\end{equation}
and the on-shell conditions $p^2_1=p^2_2=0,\ p^2_3=p^2_4=m^2_\pi$ one gets the
following constraints between the $C_i$
$$
  C_1+C_2(p_1p_2)+C_4(p_1p_3)=0, \quad C_1+C_2(p_1p_2)+C_3(p_2p_3)=0,
$$
\begin{equation}
  C_3(p_1p_2)+C_5(p_1p_3)=0, \quad C_4(p_1p_2)+C_5(p_2p_3)=0.
\label{a12}                                       
\end{equation}

Expressing all the scalar products $(p_ip_j)$ in terms of Mandelstam
variables
\begin{equation}
  s=(p_1+p_2)^2, \quad t=(p_1-p_3)^2, \quad u=(p_1-p_4)^2, \quad
  s+t+u=2m^2_\pi ,
\label{stu}                                       
\end{equation}

\noindent redefining $C_2=A(s,t,u)$ and $C_5=-sB(s,t,u)$, and
with the following useful designations,
\begin{equation}
  \xi_n=\xi -nm^2_\pi , \quad \xi =s,t,u, \quad
  Y=ut-m^4_\pi =u_1t_1-sm^2_\pi,
\label{xi}
\end{equation}

\noindent one gets

\begin{equation}
  T_{\mu \nu}=A(s,t,u){\cal L}_{1 \mu\nu}+B(s,t,u){\cal L}_{2 \mu\nu},
\label{a13}                                       
\end{equation}
with the gauge invariant tensors
\begin{equation}
  {\cal L}_1^{\mu\nu}=p_2^{\mu}p_1^{\nu}-\frac{s}{2}g^{\mu\nu},
\label{a14}                                       
\end{equation}
\begin{equation}
  {\cal L}_2^{\mu\nu}=-\left(\frac{u_1t_1}{2}g^{\mu\nu}+
  t_1p_2^\mu p_3^\nu +u_1p_3^\mu p_1^\nu +sp_3^\mu p_3^\nu\right).
\label{a14a}                                      
\end{equation}

The amplitudes $H_{++}$ ($H_{+-}$) with equal (opposite) helicity photons are
\begin{equation}
  H_{++}=-(A+m^2_\pi B), \quad H_{+-}=\frac{Y}{s}B.
\label{hel}                                       
\end{equation}
The differential cross section for $\gamma\gamma\to\pi\pi$ with averaged
photon polarizations in the center-of-mass system are calculated by the
standard expression
\begin{eqnarray}
\label{crossection}
  \frac{d\sigma^{\gamma\gamma\to\pi\pi}}{d\Omega}&=&\frac{\alpha^2s}{32{\cal
  S}}\beta (s)H(s,t), \nonumber \\
  \beta (s)&=&(1-4m^2_\pi /s)^{1/2}, \nonumber \\
  H(s,t)&=&|H_{++}|^2+|H_{+-}|^2.
\end{eqnarray}
The factor ${\cal S}$ is $1$ ($2$) for $\pi^+\pi^-$ ($\pi^0\pi^0$). Let
$\theta$ be the scattering angle in the center-of-mass system. To compare
our results with experiment, the maximum value of the integration variable
$(\cos{\theta})_{\rm max}$ is equal to $0.6$ for $\pi^+\pi^-$ \cite{5} and
$0.8$ for $\pi^0 \pi^0$ \cite{6}.

\vspace{0.5cm}
\section{The model}
\vspace{0.5cm}

{\it 3.1. The Lagrangian}
\vspace{0.5cm}

The original SU(2) flavor NJL Lagrangian \cite{12} with the fermionic
degrees of freedom reinterpreted as the light up, $u$, and down, $d$,
quarks incorporates the essential symmetries for the description of
systems involving pions. We use it here in the form
\begin{equation}
   {\cal L}=\bar{\psi}[i\!\!\not{\!\partial}
   -{\cal M}-eQ\!\not{\!\!A} ]\psi +\frac{G}{2}\left[(\bar{\psi}\psi)^{2}
  +(\bar{\psi}i\gamma_{5}\tau_{i}\psi)^{2}\right],
\label{lagr}
\end{equation}
where $G$ is the strong coupling constant of the four-fermion interaction
and ${\cal M}$ is the diagonal current quark mass matrix explicitly
breaking the chiral symmetry of the Lagrangian. We use
$\hat{m}_u=\hat{m}_d=\hat{m}$ for the matrix elements of ${\cal M}$. The
electromagnetic interactions are introduced by the replacement
$\partial_\mu\rightarrow\partial_\mu+ieQA_\mu$, where $Q=(1+3\tau_3)/6$ is
the charge matrix in the isotopic space.

Pions appear as the Goldstone modes associated with the spontaneous
breaking of chiral symmetry. The constituent quark masses ($m$) are
generated by the gap equation
\begin{equation}
m-\hat{m}=8mGI_1.
\label{gap}
\end{equation}
We give all necessary definitions for quark-loop integrals $I_i$ in the
Appendix. We use the Pauli-Villars regularization
\cite{18}-\cite{20}, introducing one Pauli-Villars regulator in such a way
\cite{17}, that the basic scalar integrals $I_i$ coincide with those of
the usual
regularization scheme with a covariant
four-momentum cutoff, $\Lambda$. In the following we shall refer to the
quark one-loops of NJL as tree contributions (in terms of meson fields).
To obtain the two-pion production
amplitude on the
basis of Lagrangian (\ref{lagr}) one can use the momentum-space
bosonization method \cite{13} or any other
\cite{14}-\cite{fmd}. Let us stress that we calculate the tree
level contributions at all orders in $p^2$, in leading $1/N_c$.
In chiral perturbation theory this corresponds to the coefficients
of the operators in the tree level Lagrangian to all orders in a
momentum expansion.

The collective variables describing meson excitations are taken in the
form of quark-antiquark field combinations linearly transformed with
respect to the chiral group action. This method is completely equivalent
to the existing nonlinear approaches but yields fewer Feynman diagrams in
concrete calculations. We already discussed this issue in detail in
\cite{17}.

\vspace{1cm}

\noindent
{\it 3.2. The $\gamma\gamma\rightarrow\pi\pi$ amplitude in the NJL model}
\vspace{0.5cm}

Our task is to calculate contributions from quark one-loop diagrams to the
amplitudes $A(s,t,u)$ and $B(s,t,u)$ within the NJL model. Before going
into detail, we make a few general comments. With the NJL model, one can
calculate arbitrary $N$-point functions to all orders in external momenta.
These calculations correspond to allowing for all tree contributions, i.e.
terms leading in $1/N_c$. The next step is to examine one-loop meson
diagrams \cite{21}. However one has to confine oneself only to the first
terms of the momentum expansion of the effective meson vertices derived
from quark loops. Now it is not clear how to do it in a general case. Thus
the following point of view seems reasonable. If there is a tree
contribution, considering meson loop diagrams results in small
corrections which can be ignored in the first approximation. And only if
there is no tree contribution for whatever reason, consideration of
meson loops is mandatory. This idea underlies for example the calculations
in \cite{11}. We shall follow it.

The process $\gamma\gamma\rightarrow\pi^+\pi^-$ is known to have tree
contributions in all orders of the chiral expansion. It is these
contributions that we shall calculate by the NJL model here. For the
neutral $\gamma\gamma\rightarrow\pi^0\pi^0$ channel, tree contributions
appear only
in the $p^6$ and higher approximations. At a level of $p^4$ the amplitudes
$A(s,t,u)$ and $B(s,t,u)$ are only determined by meson loop diagrams
calculated in \cite{1,9}. This result does not depend on the model used
and is of general character. We shall use it as a leading
approximation\footnote{We ignore the kaon loop contribution, since
we work in SU(2); its contribution is small in the kinematical
region that we address \cite{1}.}.
\begin{equation}
 A^N_{c.l.}=-4\frac{s-m^2_\pi}{sf^2}\G_\pi (s), \quad B^N_{c.l.}=0.
\label{bij}                                       
\end{equation}
Here $f$ is the value of pion decay constant $f_\pi$ in the chiral limit.
The meson
loop function $\G_\pi (s)$ is given in the Appendix\footnote{The index
$\pi$ of the function $\G_\pi (s)$ indicates that this function
differs from the function $\G (s)$ given in the Appendix only by having
the pion mass in place of the quark mass.}. We use the notation of
\cite{7} for the loop functions. Further, when describing the process, we
shall calculate quark loop corrections to the leading contribution.

Let us turn to calculations.
Contributing to the one-loop-order $\gamma\gamma\rightarrow\pi\pi$
amplitude  in the NJL model are box diagrams of quarks, with two photons
and two pions attached to the vertices, scalar exchange in the $s$-channel
and pion exchange in the $u$ and $t$ channels. These diagrams are shown in
fig.1. Pion exchange contributes only in the case of
$\gamma\gamma\rightarrow\pi^+\pi^-$.

We proceed to calculate explicitly the contribution of the box diagrams, which
is the most difficult to obtain. There are a total of six diagrams, with two
distinct classes, one with photons being neighbors (four of this kind), see
fig.1a, and the one with photons alternating with pions (two of this kind).
The complete amplitude for the box diagrams, $T^{(box)}_{\mu\nu}$, is
\begin{eqnarray}
  T^{(box)}_{\mu\nu}&=&4g_{\pi}^2Z_\pi^{-1}
    \left\{f_1^{ab}[J_{\mu\nu}^{(1)}(p_1,p_1+p_2,p_3)
                   +J_{\mu\nu}^{(1)}(p_1,p_1+p_2,p_4)]
    \right.\nonumber \\
  &+&\left. f_2^{ab}J_{\mu\nu}^{(2)}(p_1,p_1-p_3,p_4)\right\}.
\label{box}                                       
\end{eqnarray}
Here $g_{\pi}$ is the $\pi q\bar{q}$ coupling constant,
\begin{equation}
\label{gpi}
  g_\pi^2=\frac{1}{4I_2(m^2_\pi )},\quad Z_\pi =1+\frac{m^2_\pi}{I_2(m^2_\pi
  )}\frac{\partial I_2(p^2)}{\partial p^2}\bigg\vert_{p^2=m^2_\pi}.
\end{equation}
The factors $f_i^{ab}$ are the result of the
trace over flavor. They are equal to
\begin{equation}
\label{f12}
  f^{ab}_1=\frac{10}{9}\delta_{ab}, \quad f^{ab}_2=\frac{2}{9}\left(
  9\delta_{a3}\delta_{b3}-4\delta_{ab}\right).
\end{equation}
As follows from these formulae, in the channels in question one has
$f_1^N=f_2^N=f_1^C=10/9, \quad f_2^C=-8/9$,
where we use the symbols $N$ and $C$ for the neutral and charged two-pion
production respectively.

The Lorentz tensors $J_{\mu\nu}^{(i)}$ are the following two integrals,
related to two topologically different box diagrams of fig.1a
\begin{equation}
  J^{(1)}_{\mu\nu}(k_1,k_2,k_3)=\left(\frac{N_c}{4i}\right)
  \int_\Lambda\frac{d^4p}{(2\pi )^4}\mbox{Tr}
  [S(p)\gamma_\mu S(p+k_1)\gamma_\nu S(p+k_2)\gamma_5S(p+k_3)\gamma_5],
\label{j1}                                        
\end{equation}
and
$$
  J^{(2)}_{\mu\nu}(k_1,k_2,k_3)=J^{(2)}_{\nu\mu}(k_3-k_2,-k_2,k_1-k_2)=
$$
\begin{equation}
  =\left(\frac{N_c}{4i}\right)\int_\Lambda\frac{d^4p}{(2\pi )^4}\mbox{Tr}
  [S(p)\gamma_\mu S(p+k_1)\gamma_5 S(p+k_2)\gamma_\nu S(p+k_3)\gamma_5],
\label{j2}                                        
\end{equation}
where the standard notation is used,
\begin{equation}
  S(p)=\frac{\not{\! p}+m}{p^2-m^2},
\label{a24}                                       
\end{equation}
\noindent and $N_c=3$ is the number of colors.
As a consequence of Furry's theorem, the following identities hold
\begin{equation}
  J^{(1)}_{\mu\nu}(p_1,p_1+p_2,p_3)=J^{(1)}_{\nu\mu}(p_2,p_1+p_2,p_4)\;,
\label{a25}                                       
\end{equation}
\begin{equation}
  J^{(1)}_{\mu\nu}(p_1,p_1+p_2,p_4)=J^{(1)}_{\nu\mu}(p_2,p_1+p_2,p_3)\;,
\label{a26}                                       
\end{equation}
\begin{equation}
  J^{(2)}_{\mu\nu}(p_1,p_1-p_3,p_4)=J^{(2)}_{\nu\mu}(p_2,p_2-p_3,p_4)\;.
\label{a27}                                       
\end{equation}
These identities were used to obtain (\ref{box}).

One can find that
\begin{eqnarray}
  &&J^{(1)}_{\mu\nu}(p_1,p_1+p_2,p_3)+
    J^{(1)}_{\mu\nu}(p_1,p_1+p_2,p_4)=\nonumber \\
  &=&\beta_{\mu\nu}-\frac{1}{s}{\cal L}_{1\mu\nu}\kappa_2(s,t,u)
    +\frac{1}{Y}\left(m^2_\pi {\cal L}_{1\mu\nu}-{\cal L}_{2\mu\nu}
     \right)\kappa_3(s,t,u),
\label{resj1}                                     
\end{eqnarray}
$$
  J^{(2)}_{\mu\nu}(p_1,p_1-p_3,p_4)=
$$
\begin{equation}
  =-\beta_{\mu\nu}+\frac{1}{s}{\cal L}_{1\mu\nu}\kappa_4(s,t,u)
    -\frac{(s-2m^2_\pi )}{Y}\left(m^2_\pi {\cal L}_{1\mu\nu}-
    {\cal L}_{2\mu\nu}\right)\kappa_5(s,t,u).
\label{resj2}                                     
\end{equation}
Here we use the following notation
\begin{eqnarray}
  \beta^{\mu\nu}&=&\frac{1}{2}g^{\mu\nu}\kappa_1(s,t,u)+
  (p_3^\mu p_3^\nu +p_4^\mu p_4^\nu )(\Delta_tI_2+\Delta_uI_2)+
  \nonumber \\
  &+&\frac{1}{s}(s-2m^2_\pi )(p_2^\mu p_3^\nu -p_3^\mu p_1^\nu )
  (\Delta_tI_2-\Delta_uI_2), \\
  \Delta_{\xi}I_2&=&\frac{1}{\xi_1}[I_2(\xi )-I_2(m^2_\pi )],  \\
  \kappa_1(s,t,u)&=&I_2(t)+I_2(u)+\frac{1}{s}(s-2m^2_\pi )
  (u_1\Delta_tI_2+t_1\Delta_uI_2), \\
  \kappa_2(s,t,u)&=&3\tilde{Q}_3(s)+(s-2m^2_\pi )[\Delta_tI_2+\Delta_uI_2
  +I_3(p_1,p_3)+I_3(p_2,p_3)- \nonumber \\
  &-&Q_4(p_1,p_1+p_2,p_3)-Q_4(p_1,p_1+p_2,p_4)], \\
  \kappa_3(s,t,u)&=&sI_3(p_1,-p_2)+t_1I_3(p_1,p_3)+u_1I_3(p_2,p_3)
  +(s-2m^2_\pi )\kappa_6(s,t,u)- \nonumber \\
  &-&\frac{s}{2}\left[tI_4(p_1,p_1+p_2,p_3)+uI_4(p_1,p_1+p_2,p_4)\right], \\
  \kappa_4(s,t,u)&=&(s-2m^2_\pi )[\Delta_tI_2+\Delta_uI_2
  +Q_4(p_1,p_1-p_3,p_4)-I_3(p_2,p_3)]+
  \nonumber \\
  &+&\frac{2m^2_\pi}{s}\left[t_1I_3(p_1,p_3)+u_1I_3(p_2,p_3)-
  \frac{Y}{2}I_4(p_1,p_1-p_3,p_4)\right], \\
  \kappa_5(s,t,u)&=&\frac{1}{s}\left[I_2(t)+I_2(u)-2I_2(m^2_\pi )\right]
  +Q_4(p_1,p_1-p_3,p_4)-I_3(p_2,p_3), \\
  \kappa_6(s,t,u)&=&\frac{1}{s}\left[I_2(t)+I_2(u)-2I_2(m^2_\pi )\right]+
  I_3(p_3,-p_4)-Q_4(p_1,p_1+p_2,p_3)-\nonumber \\
  &-&Q_4(p_1,p_1+p_2,p_4)-\frac{1}{2Y}\left\{(s-2m^2_\pi )[2t_1I_3(p_1,p_3)+
  2u_1I_3(p_2,p_3)+\right. \nonumber \\
  &+&sI_3(p_1,-p_2)+sI_3(p_3,-p_4)]-2m^2_\pi [u_1I_3(p_1,p_3)+t_1I_3(p_2,p_3)+
  \nonumber \\
  &+&\left.
  sI_3(p_3,-p_4)]+s[t^2I_4(p_1,p_1+p_2,p_3)+u^2I_4(p_1,p_1+p_2,p_4)]
  \right\}.
\label{kappas}                                    
\end{eqnarray}
The rest of the notation is given in the Appendix. Note again that we
systematically use the conditions (\ref{a9}) when deriving (\ref{resj1})
and (\ref{resj2}).

Now let us evaluate the scalar exchange, which contributes only to the
s-channel, see fig.1b. Here one can obtain
\begin{equation}
  T^{(\sigma )}_{\mu\nu}=8g_{\pi}^2Z_\pi^{-1}f_1^{ab}{\cal L}_{1\mu\nu}
  m^2Q_3(s)\frac{[2I_2(s)+(s-2m^2_\pi )I_3(p_3,-p_4)]}
                {\left[\frac{\hat{m}}{4mG}+(4m^2-s)I_2(s)\right]}.
\label{sigma}                                     
\end{equation}
The function $Q_3(s)$ results from calculation of a triangular quark
diagram describing two-photon decay of a scalar meson. It is given in the
Appendix. The expression in the numerator is well known. It corresponds to
the $\sigma\pi\pi$ vertex. The denominator stems from the scalar particle
propagator. It is equal to $[m^2_\sigma (s)-s]/4g_\pi^2(s)$. The derived
expression (35) is gauge-invariant.

Finally, we evaluate pion exchange, $T_{\mu\nu}^{(\pi )}$, which
contributes to the $t$ and $u$ channels in the case of production of
charged pions, see fig.1c. For on-shell photons we obtain
\begin{eqnarray}
  T^{(\pi )}_{\mu\nu}&=&8g_{\pi}^2Z_\pi^{-1}(f_1^{ab}-f_2^{ab})
  \left\{\frac{p_{3\mu} (p_1-p_3)_\nu}{t_1^2}\left[tI_2(t)-m^2_\pi
         I_2(m^2_\pi )\right]+ \right. \nonumber \\
  &+&\left.
  \frac{p_{3\nu} (p_2-p_3)_\mu}{u_1^2}\left[uI_2(u)-m^2_\pi
  I_2(m^2_\pi )\right]\right\}.
\label{pion}                                      
\end{eqnarray}
In eq.(\ref{pion}) the pion propagator is cancelled by one of the
$\gamma\pi\pi$ vertices.

The sum of (\ref{box}) and (\ref{pion}) meets the general requirement of
gauge invariance because it is proportional only to the Lorentz tensors
${\cal L}_1^{\mu\nu}$ and ${\cal L}_2^{\mu\nu}$
\begin{equation}
\label{bp}
  T^{(box+\pi )}_{\mu\nu}=4g_{\pi}^2Z_\pi^{-1}[f_1^{ab}R_{1\mu\nu}+
  f_2^{ab}R_{2\mu\nu}]
\end{equation}
with
\begin{eqnarray}
  R_1^{\mu\nu}&=&2\kappa_7{\cal L}_2^{\mu\nu}-\frac{1}{s}\bar{\kappa}_2
  {\cal L}_1^{\mu\nu}+\frac{1}{Y}(m^2_\pi {\cal L}_1^{\mu\nu}-
  {\cal L}_2^{\mu\nu})\kappa_3 ,  \\
  R_2^{\mu\nu}&=&-2\kappa_7{\cal L}_2^{\mu\nu}+\frac{1}{s}\bar{\kappa}_4
  {\cal L}_1^{\mu\nu}-\frac{1}{Y}(m^2_\pi {\cal L}_1^{\mu\nu}-
  {\cal L}_2^{\mu\nu})(s-2m^2_\pi )\kappa_5 .
\label{R}                                         
\end{eqnarray}
Apart from the known functions $\kappa_i(s,t,u)$ (see
(29)-(\ref{kappas})), we also use new ones
\begin{eqnarray}
  \kappa_7&=&\frac{m^2_\pi}{s}\left[\frac{I_2(t)}{t_1^2}+\frac{I_2(u)}{u_1^2}
  \right]-\frac{I_2(m^2_\pi )}{u_1t_1}\left[1+\frac{m^2_\pi
  (s-2u_1t_1)}{st_1u_1}\right], \\
  \bar{\kappa_2}&=&\kappa_2-s(\Delta_tI_2+\Delta_uI_2), \\
  \bar{\kappa_4}&=&\kappa_4-s(\Delta_tI_2+\Delta_uI_2).
\label{newkap}                                    
\end{eqnarray}

Thus, we have all that is necessary for exploring physical consequences of
the process under consideration. Before doing this, we point out that
besides direct computation we also use another method for calculating
amplitudes, namely projection of the sought-for amplitude
$T_{\mu\nu}(s,t,u)$ by means of convolutions with tensors ${\cal
L}_1^{\mu\nu}$ and ${\cal L}_2^{\mu\nu}$.  A symbolic algebra program has
been written for this purpose in REDUCE.  First, the contributions to the
amplitudes are calculated as functions of tensor integrals for all
diagrams. Then the well known reduction method \cite{22} is applied in
order to transform the tensor integrals to a combination of a few basic
scalar integrals (\ref{a1})-(\ref{a5}). Finally, the amplitudes thus
obtained are projected onto the gauge invariant tensors (\ref{a14}) and
(\ref{a14a}). The last step is necessary in order to have few compact and
simple expressions, which are independent. This represents a consistency
check for the methods used and in particular for the regularization scheme
applied to the quark loop integrals.

\section{The $\gamma\gamma\rightarrow\pi^0\pi^0$ mode}
\vspace{0.5cm}

{\it 4.1. Chiral expansion of the amplitude}
\vspace{0.5cm}

Theoretically of greatest interest is the neutral mode. Here the leading
$p^4$ approximation is described by meson loops (see (\ref{bij})). Yet, as
was shown in the literature, this is not enough to describe the process
quantitatively even in the near-threshold region. Let us elucidate the role
of tree corrections. To this end, we isolate terms of order $p^6$ and
$p^8$ from our general result. These terms are most important near the
threshold. We start by expanding the amplitudes in a series of
momenta. For box-type diagrams we get
\begin{eqnarray}
  T^{(box)N}_{\mu\nu}&=&\frac{40g^2_\pi}{9Z_\pi}\left\{
  {\cal L}_{1\mu\nu}\left[-\frac{2}{3}I_3+\left(m^2_\pi -\frac{2s}{15}
  \right)I_4+ \right.\right.\nonumber \\
  &+&\left.\left.\left(\frac{2s^2}{7}-\frac{u_1t_1}{2}+\frac{4m^4_\pi}{3}
  \right)I_5\right]-\frac{1}{2}{\cal L}_{2\mu\nu}\left[I_4-
  \left(\frac{s-14m^2_\pi}{3}\right)I_5\right]\right\}+\ldots
\label{expb}                                      
\end{eqnarray}

\noindent where the superscript $N$ stands for neutral channel.
From here on we use the following notation
\begin{eqnarray}
  I_2&=&I_2(0)=\frac{3}{16\pi^2}\left(\ln\frac{\Lambda^2+m^2}{m^2}-
  \frac{\Lambda^2}{\Lambda^2+m^2}\right), \\
  I_3&=&I_3(0)=-\frac{3 h_1}{32\pi^2m^2},\quad
  h_1=\left(\frac{\Lambda^2}{\Lambda^2+m^2}\right)^2, \\
  I_4&=&I_4(0)=\frac{h_2}{32\pi^2m^4},\quad
  h_2=\left(\frac{\Lambda^2+3m^2}{\Lambda^2+m^2}\right)h_1, \\
  I_5&=&\frac{h_1(\Lambda^4+4\Lambda^2m^2+6m^4)}{320\pi^2m^6
       (\Lambda^2+m^2)^2}.
\label{Ih}                                        
\end{eqnarray}
From the scalar exchange we have
\begin{eqnarray}
  T^{(\sigma )N}_{\mu\nu}&=&\frac{40g^2_\pi}{9Z_\pi}{\cal L}_{1\mu\nu}
  \left[\frac{2}{3}I_3-\frac{7s}{60}I_4+\frac{s-m^2_\pi}{6m^2}I_3+
  \frac{s-2m^2_\pi}{3I_2}I_3^2+ \right.\nonumber \\
  &+&\frac{(s-m^2_\pi )^2}{24m^4}I_3+\frac{3s^2-11sm^2_\pi +8m^4_\pi}{
  36m^2I_2}I^2_3+\frac{s(s-2m^2_\pi )}{9I_2^2}I_3^3- \nonumber \\
  &-&\left.\frac{7s(s-m^2_\pi )}{240m^2}I_4+\frac{40m^4_\pi\!+\!14sm^2_\pi\!
  -\!17s^2}{120I_2}I_3I_4-\frac{5s^2}{42}I_5+\ldots\right].
\label{exps}                                      
\end{eqnarray}
As might be expected, by summing (\ref{expb}) and (\ref{exps}) the
constant terms ($2I_3/3$) are cancelled (absence of tree contributions at
a level of $p^4$), and others are grouped into a combination
\begin{eqnarray}
  T^{(box+\sigma )N}_{\mu\nu}&\!=\!&\frac{40g^2_\pi}{9Z_\pi}\left\{
  {\cal L}_{1\mu\nu}\left[\frac{(s\!-\!m^2_\pi )}{6m^2}\left(
  1+\frac{s\!-\!m^2_\pi}{4m^2}\right)I_3-\frac{1}{4}\left(s\!-\!4m^2_\pi +
  \frac{7s(s\!-\!m^2_\pi )}{60m^2}\right)I_4+ \right.\right.\nonumber \\
  &\!+\!&\left(s-2m^2_\pi +\frac{3s^2\!-\!11sm^2_\pi\!+\!8m^4_\pi}{12m^2}
  \right)\frac{I_3^2}{3I_2}+\frac{s(s\!-\!2m^2_\pi )}{9I^2_2}I_3^3+\nonumber\\
  &\!+\!&\left.\frac{40m^4_\pi\!+\!14sm^2_\pi\!-\!17s^2}{120I_2}I_3I_4+
  \frac{1}{6}\left(s^2-3u_1t_1+8m^4_\pi\right)I_5\right]- \nonumber \\
  &\!-\!&\left.\frac{1}{2}{\cal L}_{2\mu\nu}\left[I_4-\left(\frac{
  s-14m^2_\pi}{3}\right)I_5\right]\right\}+\ldots
\label{expbs}                                     
\end{eqnarray}
Now let us switch to the chiral expansion of the obtained amplitude. This
is an expansion in external momenta and masses of current quarks. We
discussed the problem in detail within the NJL model in \cite{17}, where
all necessary formulae can be found. The quantities corresponding to the
leading terms of the chiral expansion will be labelled by an ``$\circ$"
above the respective symbol (except the chiral limit for the
constant $f_\pi$ traditionally
designated by $f$). By algebraic transformations, we find (see (\ref{a13}))
for  the amplitudes $A(s,t,u)$ and $B(s,t,u)$
\begin{eqnarray}
  A^N(s,t,u)&\!=\!&\frac{40\!\mo}{9f^2}\left\{
  \frac{\Ito}{6\!\mo}\left(s-\!\mpio +\frac{(s-2\!\mpio )^2}{4\!\mo}\right)
  +\frac{\Ito^2}{3\!\Ido}\left[(s-2\!\mpio )\left(1-\frac{5\!\mpio}{6\!\mo}
  \right)+\right.\right.\nonumber \\
  &\!+\!&\left.\frac{s(s-2\!\mpio )}{4\!\mo}\right]-\frac{\Ico}{4}\left[
  (s-4\!\mpio )\left(1-\frac{3\!\mo\mpio}{(\Lambda^2+\!\mo )(\Lambda^2+3\!\mo
  )}\right)+\right. \nonumber \\
  &\!+\!&\left.\frac{7s(s-\!\mpio )}{60\!\mo}-
  \frac{2\!\mpio (s-3\!\mpio )}{\mo}\right]+\frac{\Ito^3}{9\Ido^2
  }\left(s^2-6s\!\mpio +12\!\mpioo\right)+ \nonumber \\
  &\!+\!&\left.\frac{\Ito\Ico}{120\!\Ido}\left[16\!\mpio (9s-20\!\mpio )-17s^2
  \right]+\frac{\Ipo}{6}\left(s^2-3u_1t_1+8\!\mpioo\right)\right\},  \\
\label{chexpone}                                     
  B^N(s,t,u)&\!=\!&-\frac{20\!\mo\Ico}{9f^2}\left[1-\frac{\mpio}{\mo}-
  \frac{3\!\mo\mpio\ho^2}{\Lambda^4\hoo}-\frac{\mpio\Ito}{3\!\Ido}+
  \frac{14\!\mpio\!-s}{3\!\Ico}\Ipo\right].
\label{chexptwo}                                     
\end{eqnarray}
These expressions describe the tree-level contributions to the amplitudes
$A^N(s,t,u)$ and $B^N(s,t,u)$, which involve terms of orders $p^6$ and
$p^8$. At $\Lambda \rightarrow\infty$ the term proportional to 1 in the
amplitude $B^N(s,t,u)$ coincides with the expression derived in \cite{11}.
The terms of order $p^6$ in the amplitude $A^N(s,t,u)$ differ from the
result of \cite{11}. The discrepancy is due to the contribution from the
$\sigma$ meson exchange diagram, which was not considered there.

In the $p^6$ approximation the CHPT Lagrangian has three constants $d_1,\
d_2$ and $d_3$ entering in the description of the process
$\gamma\gamma\rightarrow\pi^0\pi^0$ \cite{7}. In terms of $d_i$ the
amplitude of order $p^6$ in the process in question has the form
\begin{eqnarray}
  A^{(6)}(s,t,u)&=&-\frac{40}{9f^4}\left[(s-2\!\mpio )(4d_2-d_1-8d_3)+
  2(s-\!\mpio )(d_1+4d_3)\right], \\
  B^{(6)}(s,t,u)&=&\frac{80d_1}{9f^4}.
\label{Gasser}
\end{eqnarray}
It is easy to see that the NJL model yields the following values for the
constants
\begin{eqnarray}
  d_1&=&-\frac{2f^2\!\hoo}{(16\pi\!\m )^2}=-1.26\times 10^{-4}, \nonumber \\
  d_2&=&\frac{f^2}{(16\pi\!\m )^2}\left[\!\ho +\!\hoo -\frac{3\!\mo\ho^2}{
  (2\pi f)^2}\right]=0.96\times 10^{-4}, \nonumber \\
  d_3&=&\frac{f^2\!\ho}{2(16\pi\!\m )^2}=2.89\times 10^{-5}.
\label{di}
\end{eqnarray}
The analytical expressions for these coefficients fully agree with the
corresponding expressions derived in \cite{bb}, if one sets $g_A=1$ there.
For quantitative evaluation we used the numerical values of the model
parameters fixed earlier in \cite{13}:  $\m =221.2\ \mbox{MeV},\ f=88.6\
\mbox{MeV},\ \mpi =141.5\ \mbox{MeV}$ and $\Lambda =1\ \mbox{GeV}$. Note
that these values correspond to the leading terms of the chiral expansion
of the physical quantities chosen as follows: $m_\pi =139\ \mbox{MeV},\ f_\pi
=93.1\ \mbox{MeV},\ m=241.8\ \mbox{MeV},\ \Lambda =1\ \mbox{GeV}$.

An equivalent $d_i$-related set of parameters is also used. It includes
constants $a_1,\ a_2$ and $b$.
\begin{eqnarray}
  \frac{a_1}{(16\pi^2f^2)^2}&=&
  \left(\frac{20}{9f^4}\right)16(d_3-d_2),\nonumber \\
  \frac{a_2}{(16\pi^2f^2)^2}&=&
  \left(\frac{20}{9f^4}\right)(d_1+8d_2),\nonumber  \\
  \frac{b}{(16\pi^2f^2)^2}&=&-\left(\frac{10}{9f^4}\right)d_1.
\label{ab}
\end{eqnarray}
The values we obtain for them are
\begin{eqnarray}
  a_1&=&\frac{10(4\pi f)^2}{9\!\mo}\left[\frac{3\!\mo\ho^2}{2\pi^2f^2}-\!\ho
  -2\!\hoo\right]=-59.5  \nonumber \\
  a_2&=&\frac{10(2\pi f)^2}{9\!\mo}\left[4\!\ho +3\!\hoo -
  \frac{3\!\mo\ho^2}{\pi^2f^2}\right]=35.6  \nonumber \\
  b&=&\frac{5(2\pi f)^2}{9\!\mo}\hoo =3.5
\label{an}
\end{eqnarray}
Let us compare these values with the results of \cite{bfp}, where the
parameters $a_1, a_2$ and $b$ were calculated within the extended NJL
model.
\begin{eqnarray}
 a_1^{(ENJL)}&=&-23.3, \nonumber \\
 a_2^{(ENJL)}&=& 14.0, \nonumber \\
   b^{(ENJL)}&=& 1.66.
\label{bi}
\end{eqnarray}
It is quite simple to account for the discrepancy. In the extended NJL
model $a_1, a_2$ contain a factor $g_A^2\simeq 0.37$ and $b$ contains a
factor $g_A\simeq 0.61$. This factor arises from allowing for vector
mesons. In the NJL without vector mesons it is equal to 1. This is the
reason for larger values of the parameters in (\ref{an}). Yet, this does
not mean that the NJL model without vector mesons cannot be used to
describe the process in question qualitatively. This only indicates that
chiral expansions in the NJL model converge slower than in the ENJL model,
which we already know from \cite{17}. Really, this means that if we want
to compare the NJL result with the CHPT calculations, where chiral
expansions converge quite fast, we must compare the {\it total} NJL
result, which takes into account all orders of the chiral expansion. We
shall return to this matter again when discussing numerical values of pion
polarizabilities.
\vspace{0.5cm}

\noindent
{\it 4.2. Polarizability of neutral pions}
\vspace{0.5cm}

To obtain the pion polarizabilities we go to the crossed channel, Compton
photon-pion scattering. This is easily done by trivial replacement of
variables. The general expressions for the polarizabilities are obtained as
\begin{equation}
\label{pola1}
  \beta_{\pi}^{i}=\frac{\alpha A^{i}}{2m_\pi}\bigg\vert_{s,t_1,u_1=0}
\end{equation}
\begin{equation}
\label{pola2}
  (\alpha_{\pi} +\beta_{\pi} )^{i} =-\alpha m_\pi B^{i}\vert_{s,t_1,u_1=0}
\end{equation}
with ($i=N,C$) and the coupling constant $\alpha=e^2/(4\pi)=1/137$. In the
case of charged pions, $(C)$ calculated below in {\it 5.2}, the Born term is
removed from the amplitudes. Different phase conventions are used in
the literature. We adopt the one chosen in \cite{7}. This is the
Condon--Shortley phase convention. The one-loop result is
\begin{equation}
  (\alpha_{\pi} +\beta_{\pi} )^N=0,
\end{equation}
\begin{equation}
  (\alpha_{\pi} -\beta_{\pi} )^N=-\frac{\alpha}{48\pi^2f^2\!\mpi}
  =-1.0.
\label{pl}
\end{equation}
From here on we express the numerical values of polarizabilities
in units $10^{-4}\ \mbox{fm}^{3}$. To this $p^4$ result one must add
the $p^6$ and $p^8$  tree contributions obtained by us.
\begin{eqnarray}
  (\alpha_{\pi} +\beta_{\pi} )^N&=&
  \frac{10\alpha\!\mpi}{(12\pi f\!\m )^2}\left\{\hoo
  \left[1-\frac{79\!\mpio}{120\!\mo}-\frac{3\!\mo\mpio}{(\Lambda^2
  +\!\mo )(\Lambda^2+3\!\mo )}\right]+\right. \nonumber\\
  &+&\left.\frac{\mpio\ho\hoo}{(2\pi f)^2}+\frac{7\!\mo\mpio\ho}{5(\Lambda^2
  +\mo )^2}\right\}=1.14 \\
  \beta_{\pi}^N&=&\frac{5\alpha\!\mpi}{(12\pi f\!\m )^2}\left\{\ho\left(1-
  \frac{7\!\mpio}{8\!\mo}\right)-\frac{3\!\mo\ho^2}{2\pi^2f^2}\left(
  1-\frac{5\!\mpio}{8\!\mo}\right)+\right.\nonumber \\
  &+&2\!\hoo\left[1-\frac{11\!\mpio}{8\!\mo}-\frac{3\!\mo\mpio}{(\Lambda^2
  +\!\mo )(\Lambda^2+3\!\mo )}\right]-\frac{15\mpio\mo\ho^3}{(2\pi f)^4}+
  \nonumber\\
  &+&\left.\frac{7\!\mpio\ho\hoo}{(2\pi f)^2}+\frac{4\!\mpio\ho
  (\Lambda^4+4\Lambda^2\mo +6\!\moo )}{15\!\mo (\Lambda^2+\!\mo )^2}
  \right\}=0.92
\label{tree}
\end{eqnarray}
The numerical evaluation results are given in Table 1. It is evident from the
table how important it is to calculate polarizabilities of neutral pions with
allowance made for {\em all} terms of the chiral series and not only for its
first terms. This can be most clearly demonstrated by examining the sum and
the difference of polarizabilities. The pion loop is known not to contribute
to the sum $(\alpha_\pi +\beta_\pi )$. In other words, this quantity stems
only from many-loop contributions (meson two-loop, three-loop contributions,
etc.,).
\begin{table}
{\small {\bf Table 1:} Neutral pion polarizabilities in the NJL model in
units of $10^{-4}\ \mbox{fm}^{3}$.
The results are compared with CHPT.}\\[0.5cm]
\begin{tabular}{|c||c|c|c|c||c||c|} \hline
  & 1 loop & \multicolumn{3}{|c||}{tree}& total & CHPT \\ \cline{3-5}
  &        & $p^6$ & $p^8$ & all others & & [7] \\ \hline
 $(\alpha_{\pi} +\beta_{\pi} )^N$
  &0&$1.44$&$-0.30$&$0.05$&$1.19$&$1.17\pm 0.30$\\ \hline
 $(\alpha_{\pi} -\beta_{\pi} )^N$
  &$-1.0$&$-1.63$&$0.92$&$-0.45$&$-2.16$&$-1.90\pm 0.20$
 \\ \hline
 $\alpha_{\pi}^N$
  &$-0.5$&$-0.09$&$0.31$&$-0.20$&$-0.48$&$-0.35\pm 0.10$ \\ \hline
 $\beta_{\pi}^N$
  &$0.5$&$1.54$&$-0.62$&$0.25$&$1.67$&$1.50\pm 0.20$     \\ \hline
\end{tabular}
\end{table}
\begin{equation}
 (\alpha_\pi +\beta_\pi )^N_{CHPT}=\sum^{\infty}_{n=2}S_{(n)}=
 \frac{8\alpha\!\mpi}{(4\pi f)^4}\sum^{\infty}_{n=2}s_{(n)},
\label{s}
\end{equation}
where the index $n$ denotes a contribution from $n$-loop meson diagrams.
The quantity $s_{(n)}$ is a sum of the contact contribution $s^r_{(n)}$
($s^r_{(2)}=b^r=h^r_-$ in terms of notation from \cite{7}) and the proper
loop contribution $s^{(loops)}_{(n)}$:
\begin{equation}
 s_{(n)}=s^r_{(n)}+s^{(loops)}_{(n)}.
\label{dif}
\end{equation}
In \cite{7} they calculated the first term of (\ref{s})
$S_{(2)}=1.0+0.17$. As in (\ref{dif}), here the first term corresponds to
the contact contribution and the second term to the loop contribution. The
total tree contribution, which is calculated in the NJL model, corresponds
to the sum of the contact terms in a similar CHPT calculation. For the sum
of polarizabilities of neutral pions we obtain the value
\begin{equation}
 (\alpha_\pi +\beta_\pi )^N_{NJL}=1.19\ \sim\ \sum^{\infty}_{n=2}S^r_{(n)}.
\label{apb}
\end{equation}
Hence it can be concluded that the resonance saturation hypothesis, which
is used to determine the contact contributions in CHPT and results, among
other things in $S^r_{(2)}=1.0$, holds good here because it leads, as it
must, to fast convergence of the CHPT chiral series. It follows from our
analysis that the sum of all other terms of this series accounts for only
20\% of the leading contribution. Unlike the CHPT, where already the first
term of the chiral series practically determines the entire result, the
NJL model without vector mesons allows this at the second step. Moreover,
if we confine ourselves to the first two terms of the chiral series in
calculation of pion polarizabilities by the NJL model, we shall {\it
fully} reproduce the picture we have in the CHPT with $a^r_1$ and $b^r$ fixed
in conformity with the resonance saturation hypothesis. The parameter $a^r_2$
is not associated with the pion polarizability, so we cannot get any more
severe constraints on its value as compared with the already known ENJL result.

An interesting picture emerges for the difference of polarizabilities
$(\alpha_\pi-\beta_\pi)$. Here we have
\begin{equation}
 (\alpha_\pi -\beta_\pi )^N_{CHPT}=\sum^{\infty}_{n=1}D_{(n)}=
 \frac{8\alpha\!\mpi}{(4\pi f)^4}\sum^{\infty}_{n=1}d_{(n)}.
\end{equation}
As in (\ref{s}), the quantity $d_{(n)}$ is a sum of the contact
contribution $d^r_{(n)}$ ($d^r_{(2)}=a^r_1+8b^r=h^r_+$ in terms of
notation from \cite{7}) and the proper loop contribution
$d^{(loops)}_{(n)}$. In \cite{7} they obtained $D^r_{(2)}=-0.58$ and
$D^{(loops)}_{(2)}=-0.31$. In the NJL model, if we again confine ourselves
to the sum of the first two terms of the chiral series, we get a close
value $[(D^r_{(2)})+(D^r_{(3)})]_{NJL}=-0.70\sim (D^r_{(2)})_{CHPT}$.
However it is evident from Table 1 that the chiral series of the NJL model
for the difference of polarizabilities $(\alpha_\pi -\beta_\pi )$ converges
noticeably worse. In this case we get
\begin{equation}
 (\alpha_\pi -\beta_\pi )^N_{NJL}=-1.16\ \sim\ \sum^{\infty}_{n=2}D^r_{(n)},
\label{amb}
\end{equation}
which is twice as large (in absolute value) as the leading $(D^r_{(2)})$
result of the CHPT. Hence it follows that in this case the resonance
saturation hypothesis is less successful. This is no surprise. The
dispersion calculation methods leading to the sum rule for $(\alpha_\pi
+\beta_\pi )$ are known to yield quite an exact result for this quantity.
This is not the case with the difference $(\alpha_\pi -\beta_\pi )$ which
features uncertainties in determination of the asymptotic contribution
\cite{Petr}. It is in these cases that the complete NJL model calculations
are of particular interest. As we have just shown, they allow quite concrete
conclusions to be drawn about the efficiency of the approximations used.

Let us also stress that the correction to the electric polarizability
$\alpha_{\pi}^N$ at order $p^6$ is very small owing to strong
compensation of the terms related to $\sigma$ exchange (see the expression
in parentheses)
\begin{equation}
  \Delta^{(p^6)}\alpha_{\pi}^N=
  \frac{5\alpha\!\mpi\ho}{(12\pi f\!\m )^2}\left(
  \frac{3\!\mo\ho}{2\pi^2 f^2}-1\right)=-0.09.
\end{equation}
Therefore $p^8$ and higher order terms turn out to be essential here.
Our calculations show that the  contribution of these terms compensates almost
fully the $p^6$ result. It indicates that the two-loop meson graphs
are important for this value.

The process $\gamma\gamma\to\pi^0\pi^0$ has been used to obtain information
on the polarizabilities. We give here the results obtained from the
Crystal Ball data by Kaloshin, Persikov and Serebryakov.
$$
(\alpha_{\pi} +\beta_{\pi} )^N=1.14\pm 0.08\pm 0.16,\ \cite{kal}
$$
$$
(\alpha_{\pi} -\beta_{\pi} )^N=-1.1\pm 1.7.\ \cite{10}
$$

\vspace{0.5cm}

\noindent
{\it 4.3. Interaction cross section}
\vspace{0.5cm}

Before showing the results, we need to explain the parameters used. The
values of the parameters $m$, $f_\pi$, $m_\pi$ change at each order of the
chiral expansion, but can always be expressed in terms of the leading chiral
ones, $\m , f, \mpi$  \cite{17}, which has been done in expressions
(\ref{chexpone})-(\ref{chexptwo})
and in  equations (\ref{expchone})-(\ref{expchtwo}) below.
The Mandelstam variables are always constrained by $s+t_1+u_1=0$ and
since the pion mass changes order by order, so does also the kinematically
allowed region for $s,t$ and $u$. For that reason the pion mass which appears
in the polarization tensors ${\cal L}_1^{\mu\nu}$ and ${\cal L}_2^{\mu\nu}$
is the one calculated at a specific order. The same is true also for the
phase space factor (\ref{crossection}).
In this way we get the same result by summing an
infinite series of the chiral expansion or calculating the full amplitude.
The pion loop has been calculated only at leading order and we have to take
the pion mass of the specific tree order to which it is added.
In Fig. 2 we display the data for the cross section $\sigma (s, |\cos\theta |
\leq 0.8)$ as determined by the Crystal Ball collaboration \cite{6}. They are
shown as a function of the center-of-mass energy $W=\sqrt{s}$. The dotted
line is a well-known result arising from consideration of the leading pion
loop (\ref{bij}). The dot-and-dash line is the result at which we arrive on
taking into account $p^6$ terms. The dashed line shows the data derived from
the previous curve by including order $p^8$
corrections (\ref{chexpone})-(\ref{chexptwo}).
The solid line shows the full result. The tree contributions are clearly
seen to be responsible for enhancement of the scattering cross section in the
near-threshold region, the tree $p^6$ approximation accounting for almost the
entire effect. Diagrams with $\sigma$ exchange dominate among the diagrams
contributing to this result. Consideration of higher-order tree
corrections practically do not affect the cross section in the low-energy
region. The different thresholds correspond to the different values of the
pion mass at a certain order of the chiral expansion.

\section{The $\gamma\gamma\rightarrow\pi^+\pi^-$ mode}
\vspace{0.5cm}

{\it 5.1. The result up to $p^4$}
\vspace{0.5cm}

Let us begin with the leading-order result. Here the NJL model yields
\begin{equation}
\label{chcl}
  A^C=-\frac{\ho}{(2\pi f)^2}+...,
  \qquad B^C=-\frac{4}{u_1t_1}+...
\end{equation}
The ellipses indicate higher contributions. The amplitude resulting from
insertion of $B^C$ in (\ref{a13}) is the Born term and of order $p^2$. The
term $A^C$ leads to the $p^4$ correction of the amplitude.
Our task is to calculate
corrections to these expressions arising from consideration of NJL tree
diagrams. They can be derived from the general result given in Sec. 3.
Here we shall dwell upon the role of the $p^6$ and $p^8$ corrections which
account for the major part of the additional contribution.

\vspace{0.5cm}

\noindent
{\it 5.2. The result up to $p^6$ and higher orders in the NJL model}
\vspace{0.5cm}

Let us examine corrections to the leading terms of the chiral expansion of
the $\gamma\gamma\rightarrow\pi^+\pi^-$ amplitude. Here we have
\begin{eqnarray}
\label{expchone}
  A^{C}(s,t,u)&=&\frac{8\!\mo}{3f^2}\Ito\left\{1+\frac{5(s-\!\mpio )}{18\!\mo}
  -\frac{\mpio}{2\!\mo}\left(\frac{\Lambda^2+3\!\mo}{\Lambda^2+\mo}\right)+
  \right.\nonumber  \\
  &+&\left.\frac{\Ito}{9\!\Ido}(5s-13\!\mpio )+\frac{\Ico}{6\!\Ito}
  \left(\mpio -\frac{7s}{10}\right)\right\},\\
\label{expchtwo}
  B^{C}(s,t,u)&=&-\left\{\frac{4}{u_1t_1}\left[1+\frac{1}{2}
  \left(\frac{3\!\mpio\ho}{8\pi^2 f^2}\right)^2\right]
  +\frac{7\!\hoo}{10(6\pi f\!\m )^2}\right\}.
\end{eqnarray}

One obtains the polarizabilities for the charged pions from equations
(\ref{pola1}) and (\ref{pola2}) but with the Born term removed from
the amplitude. We derive the $p^4$ and $p^6$ orders to the polarizabilities
of charged pions to be
\begin{equation}
\label{psb}
  \beta_{\pi}^C=-\frac{2\alpha\!\ho}{\mpi\!(4\pi f)^2}\left\{
  1-\frac{5\!\mpio}{18\!\mo}\left[\frac{11}{20}+2\left(\frac{
  \Lambda^2+3\!\mo}{\Lambda^2+\!\mo}\right)-
  \frac{3\!\mo\ho}{2\pi^2f^2}\right]\right\}=-4.57,
\end{equation}
\begin{equation}
\label{psab}
  (\alpha_{\pi} +\beta_{\pi} )^C=
  \frac{7\alpha\mpi\hoo}{10(6\pi f\!\m )^2}=0.41.
\end{equation}
Generally, it is these contributions that determine the amount of the
electric and magnetic polarizabilities of charged pions. Table 2 presents
numerical evaluation results for the polarizabilities. We also included
corrections from $p^8$ terms in the table. Analytical expressions are
quite cumbersome, so we do not show the whole of them. An exception is the
sum $(\alpha_{\pi} +\beta_{\pi} )^C$,
for which the order $p^6$ contribution is a
leading one and the first correction at order $p^8$ amounts to 25\%.  The
NJL model yields for it
\begin{equation}
\label{alpb}
  (\alpha_{\pi} +\beta_{\pi} )^C=
  \frac{7\alpha\mpi\hoo}{10(6\pi f\!\m )^2}
  \left\{1-\frac{125\!\mpio}{168\!\mo}+\frac{\mpio\ho}{(2\pi f)^2}
  -\frac{13\!\mo\mpio\ho^2}{7\Lambda^4\hoo}
  \right\}=0.31.
\end{equation}
Apart from the known results obtained in Serpukhov \cite{ant} from the
reaction $\pi^-A\to\pi^-\gamma A$, the experimental body of the table
incorporates the results of analyzing angular distributions of pions in
the reaction $\gamma\gamma\to\pi^+\pi^-$ (CELLO and MARK-II data) within
the framework of the unitary model for helicity 2 amplitude \cite{kal}. As
seen from Table 2, the results of numerical calculations agree with the
available experimental data. The electric and magnetic polarizabilities
are mainly determined already at order $p^4$. Only the sum of these
quantities, which equals zero at this order, is determined by higher $p^6$
and $p^8$ corrections.
\begin{table}
{\small {\bf Table 2:} The charged pion polarizabilities in the NJL model in
units of $10^{-4}\ \mbox{fm}^{3}$. The results are compared with
phenomenological data.}\\[0.5cm]
\begin{tabular}{|c||c|c|c|c||c||c|} \hline
  & \multicolumn{4}{|c||}{tree}& total & exp. \\ \cline{2-5}
  & $p^4$ & $p^6$ & $p^8$ & all others &  &   \\ \hline
 $(\alpha_{\pi} +\beta_{\pi} )^C$&0&$0.41$&$-0.10$&$0.02$&$0.33$&
 $1.4\pm 3.1\pm 2.5$ \cite{ant}    \\
 &&&&&&$0.33\pm 0.06\pm 0.01$ \cite{kal} \\
 &&&&&&$0.22\pm 0.07\pm 0.04$ \cite{kal} \\ \hline
 $(\alpha_{\pi} -\beta_{\pi} )^C$
 &$11.62$&$-2.07$&$0.82$&$-0.36$&$10.0$&
 $15.6\pm 6.4\pm 4.4$ \cite{ant} \\
 &&&&&&$4.8\pm 1.0$  \cite{10} \\ \hline
 $\alpha_{\pi}^C$&$5.81$&$-0.83$&$0.36$&$-0.17$&$5.17$&    \\ \hline
 $\beta_{\pi}^C$ &$-5.81$&$1.24$&$-0.46$&$0.19$&$-4.84$&       \\ \hline
\end{tabular}
\end{table}

\vspace{0.5cm}

\noindent
{\it 5.3. The amplitudes: numerical results}
\vspace{0.5cm}

Let us turn to the diagrams illustrating the general formulae derived earlier.
Figure 3 displays the charged pion production cross section as a
function of energy in the center-of-mass system of interacting particles.
Integration over the scattering angle covers the interval
$|\cos\theta|<0.6$. We show four curves. The dotted curve corresponds to
the Born term derived from $B^C$ in formula (\ref{chcl}). Taking into
account corrections up to $p^4$, (\ref{chcl}) and up to $p^6$
(see formulae in (\ref{expchone})-(\ref{expchtwo})) we get the
dot-and-dash  and the dashed curves respectively.
Full calculations are represented by the solid curve. As was
expected, the cross section for production of charged pion pairs features
some enhancement caused by $p^6$ terms. The total result is close to $p^6$
estimations, in the low energy range. For higher values of the
energy the scalar resonance starts to show up in the higher orders in
momenta and in the full result.\footnote{This behavior differs from the result
obtained in \cite{wow}, where the scalar resonance was hidden. This
difference is a consequence of the different regularization schemes
used. In order to have the right low energy behavior in the
Compton photon-pion processes
considered here one has to use Pauli-Villars regularization.}

\vspace{0.5cm}
\section{Conclusions}
\vspace{0.5cm}

To summarize, we have calculated the quark one-loop contributions to the
$\gamma\gamma\rightarrow\pi\pi$ processes, using the SU(2) flavor NJL
model, with scalar-isoscalar and pseudoscalar-isovector four-quark
interactions. These calculations correspond to full consideration of tree
contributions in the examination of the above processes. We present the
amplitudes in a compact and analytical form which is used to compute the
associated cross sections and the polarizabilities of pions.

The chiral expansion of the $\gamma\gamma\to\pi^0\pi^0$ amplitude starts with
$p^4$ terms. There are no tree contributions in this approximation,
therefore the leading contribution is that from one-loop pion diagrams. We
calculated the highest tree corrections appearing at order $p^6$ and
showed that in the near-threshold region the cross section for this
process can be satisfactorily described by one-loop pion diagrams (the $p^4$
order) with the tree $p^6$ contribution. The conclusion as to the smallness of
the highest contributions (beginning with $p^8$ terms) to the cross
section of the process under study in its near-threshold region is of
interest to us. It is this kind of evaluation that the quark models are
well suited for. For example, careful examination of this problem within
the CHPT at order $p^8$ alone would require complicated calculations of
the three-loop meson diagrams.

The NJL model predicts some enhancement of the cross section for
$\gamma\gamma\to\pi^+\pi^-$ near threshold. This fact is in good
agreement with the known CHPT result \cite{1}. The increase of the
cross sections in the full result for higher energies is an artifact
of the model. This is why we have plotted them only for $W<0.4\
\mbox{GeV}$. One has to stay below the two constituent quark
threshold where the lack of confinement in the NJL model starts dominating
the result. Near the threshold this effect is absent and allows for
extraction of the polarizabilities at {\it all} tree orders.

We show that to describe the cross sections well it is enough to consider
only the first terms in a chiral expansion. Higher orders are important for
polarizabilities. Therefore one can arrive at the general conclusion that the
cross sections are very little sensitive to the polarizability. The same
conclusion has been drawn in \cite{9} in the framework of a dispersive
approach.
Among new interesting results are also calculations of the sum of the
electric and magnetic polarizabilities presented here both for neutral and
charged pions. At the one-loop meson order they are equal to zero, that is
why it is very important to advance to higher $\sim p^8$ orders. This is
possible with the NJL model, as we have shown here. \\

\vspace{0.5cm}

\noindent
{\bf Appendix. The main loop integrals}
\vspace{0.5cm}

List of the main quark-loop integrals
\begin{equation}
  I_1=iN_c\int _{\Lambda} \frac{d^4k}{(2\pi )^4} \Delta (k),
\label{a1}                                        
\end{equation}
\begin{equation}
  I_2(q_1)=-iN_c\int _{\Lambda} \frac{d^4k}{(2\pi )^4}\Delta (k)\Delta
  (k+q_1),
\label{a2}                                        
\end{equation}
\begin{equation}
  I_3(q_1,q_2)=-iN_c\int_{\Lambda}\frac{d^4k}{(2\pi )^4}\Delta (k)\Delta
  (k+q_1)\Delta (k+q_2),
\label{a3}                                        
\end{equation}
\begin{equation}
  I_4(q_1,q_2,q_3)=-iN_c\int_{\Lambda}\frac{d^4k}{(2\pi )^4}\Delta (k)\Delta
  (k+q_1)\Delta (k+q_2)\Delta (k+q_3),
\label{a4}                                        
\end{equation}
\begin{equation}
  Q_4(q_1,q_2,q_3)=-iN_c\int_{\Lambda}\frac{d^4k}{(2\pi )^4}k^2\Delta (k)
  \Delta (k+q_1)\Delta (k+q_2)\Delta (k+q_3).
\label{a5}                                        
\end{equation}
Here the notation
\begin{equation}
  \Delta(p)=\frac{1}{p^2-m^2}
\label{a6}                                        
\end{equation}
has been used. A sharp euclidean cutoff $\Lambda$ is introduced for the
scalar integrals $I_i, Q_4$, and all the cutoff dependence of
the amplitude $T_{\mu\nu}$ resides in these integrals.

We also use the standard notations for the loop functions $\bar{G}(s),
\bar{G}_\Lambda (s), \bar{J}_\Lambda (s)$ \cite{7}. They are analytic in the
complex s-plane, cut along the positive real axis for $\mbox{Re} (s)\geq 4m^2$
and $\mbox{Re} (s)\geq 4(\Lambda^2+m^2)$ respectively. At small s,
\begin{equation}
  \bar{G}(s)=\frac{1}{16\pi^2}\sum^{\infty}_{n=1}\left(\frac{s}{m^2}\right)^n
  \frac{(n!)^2}{(n+1)(2n+1)!}
\label{a7}                                        
\end{equation}
\begin{equation}
  \bar{G}_\Lambda (s)=\frac{1}{16\pi^2}\sum^{\infty}_{n=1}\left(
  \frac{s}{\Lambda^2+m^2}\right)^n
  \frac{(n!)^2}{(n+1)(2n+1)!}
\label{ag}                                        
\end{equation}
\begin{equation}
  \bar{J}_\Lambda (s)=\frac{1}{16\pi^2}\sum^{\infty}_{n=1}\left(
  \frac{s}{\Lambda^2+m^2}\right)^n
  \frac{(n!)^2}{n(2n+1)!}.
\label{aj}                                        
\end{equation}
Besides, we shall need the function
\begin{equation}
  Q_3(s)=I_3(p_1, -p_2)+\frac{N_c}{s}\tilde{Q}_3(s),
\label{qthree}                                    
\end{equation}
where
\begin{equation}
  \tilde{Q}_3(s)=\frac{4\Lambda^2}{4(\Lambda^2+m^2)-s}
  \left[\bar{J}_{\Lambda}(s)-\frac{s}{32\pi^2(\Lambda^2+m^2)}\right]
  +2\left[\bar{G}(s)-\frac{m^2}{\Lambda^2+m^2}\bar{G}_\Lambda (s)\right].
\label{aq}                                        
\end{equation}

\newpage

{\bf Figure captions}
\vspace{0.5cm}

Fig.1 - Representative Feynman diagrams involved in the calculation of
the amplitudes for $\gamma\gamma\rightarrow\pi\pi$, subject to permutations.
a) Two photons (wiggled lines) and two pions (dashed lines) are directly
coupled to the quarks (full lines). b) Quarks rescatter with the quantum
numbers of a scalar-isoscalar, $\sigma$, in the s-channel. c) When a pair of
charged pions is created, quarks can also rescatter in t- and u- channels
to form an intermediate state with the quantum numbers of a pion.
\vspace{0.5cm}

Fig.2 - The NJL $\gamma\gamma\rightarrow\pi^0\pi^0$ cross section, with
averaged photon polarizations, in comparison with the Crystal Ball data
\cite{6}. The dotted line results from the meson one-loop
calculations (the $p^4$ order). The dash-dotted line includes up to $p^6$ tree
level NJL corrections to the leading $p^4$ result. The dashed line takes into
account up to $p^8$ tree order NJL corrections and the full line corresponds
to the total tree order corrections of NJL to the pion loop.
\vspace{0.5cm}

Fig.3 - The NJL cross section for the $\gamma\gamma\rightarrow\pi^+\pi^-$
mode, with averaged photon polarizations. The dotted line corresponds to the
Born approximation ($p^2$ order), the dash-and-dot line up to $p^4$, the
dashed line up to $p^6$ order. The full line denotes the complete tree order
NJL result.

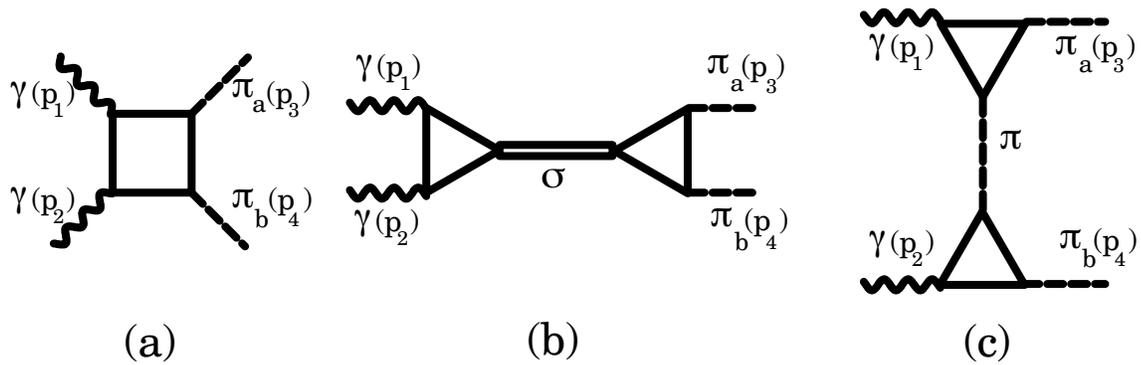
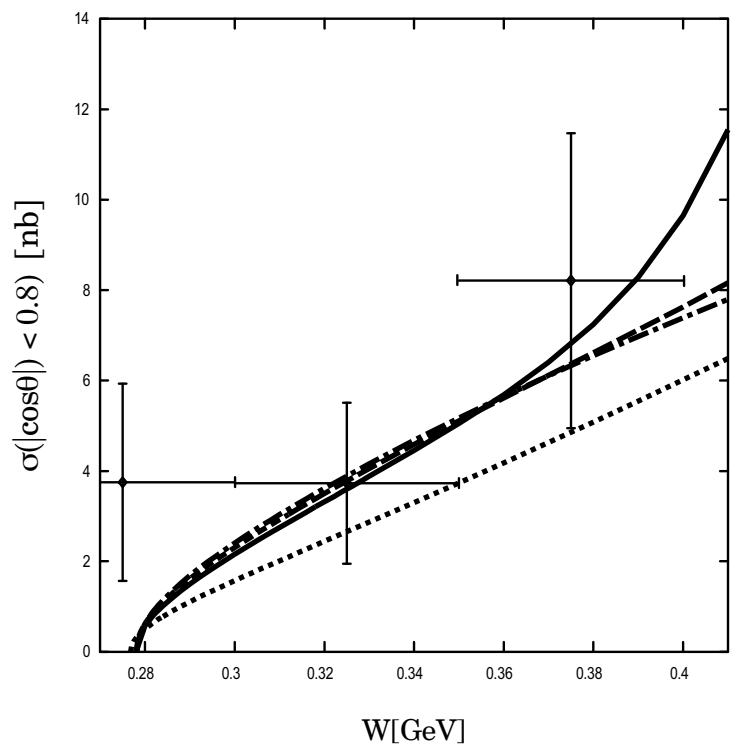
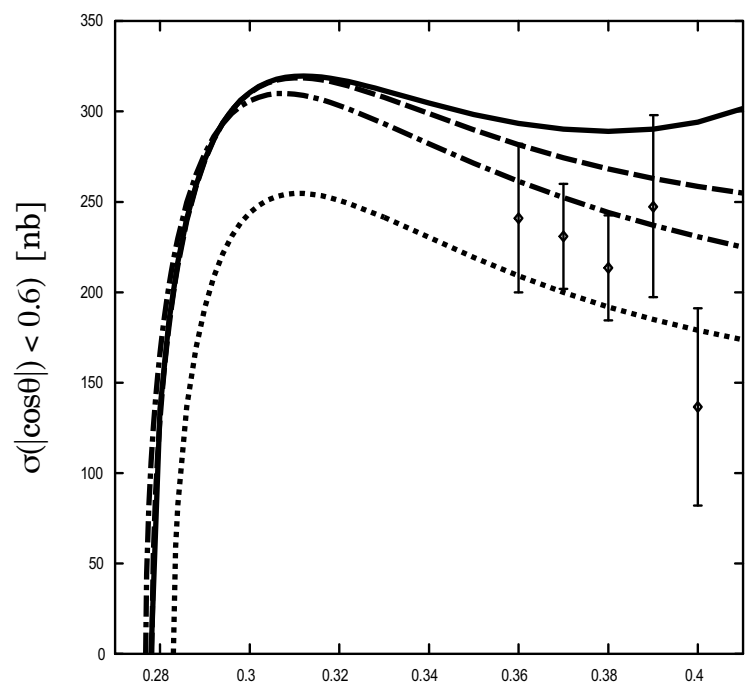
\end{document}